\documentclass[%
 aip,
 amsmath,amssymb,
 reprint,%
]{revtex4-1}

\usepackage{graphicx}
\usepackage{dcolumn}
\usepackage{bm}

\usepackage[utf8]{inputenc}
\usepackage[T1]{fontenc}
\usepackage{mathptmx}
\usepackage{etoolbox}
\usepackage{xcolor}

\usepackage[sectionbib]{bibunits}

\defaultbibliography{aipsamp}
\usepackage{standalone}
\makeatletter
\def\@email#1#2{%
 \endgroup
 \patchcmd{\titleblock@produce}
  {\frontmatter@RRAPformat}
  {\frontmatter@RRAPformat{\produce@RRAP{*#1\href{mailto:#2}{#2}}}\frontmatter@RRAPformat}
  {}{}
}%
\makeatother
\begin{document}

\preprint{AIP/123-QED}

\title{Levitation of a YIG sphere using a magnetic Paul trap - towards strongly coupled quantum magno-mechanics}
\author{A. O. Yakymenko}
\email{an.yakymenko@oist.jp}
\author{S. Das}
\author{J. Twamley}
 \affiliation{ 
   Quantum Machines Unit, Okinawa Institute of Science and Technology, 1919-1 Tancha, Onna-son, Kunigami-gun, Okinawa, Japan 904-0495
}%

\date{\today}

\begin{abstract}
Magnetic levitation offer passive levitation of massive objects for use in advanced inertia sensors,  for the generation of non-classical macroscopic motional states, and towards the table-top testing of  low energy gravity with quantum mechanics. Magnons, a quanta of spin wave, couple to many physical quantities and strongly to electromagnetic fields, even at room temperature. In this work we demonstrate the stable trapping of a small YIG sphere using a magnetic Paul trap. We present a classical stability analysis of a magnetic Paul trap and show the stability diagram for all mechanical degrees of freedom. We show that coupling between librational and translational modes changes the stability region. We experimentally levitate the soft magnet yttrium iron garnet at room temperature obtaining Q-factors of $\sim25$ and secular frequencies $15.8$ Hz and $17.2$ Hz. We provide numerical estimates of the achievable enhanced coupling between the center-of-mass motion and excited magnon modes, with a cooperativity above unity despite strong mechanical damping, indicating potential applications in quantum information processing, quantum interconnects and quantum memories.
\end{abstract}

\maketitle


\begin{bibunit}
Magnons are an example of a room temperature quantum system that has attracted significant attention over the past decade due to their ability to couple to many other types of quantum systems.
Magnons were initially formulated by Bloch in 1939, who postulated their existence as a mechanism involved in the thermal demagnetization of magnets \cite{Bloch1930ZurTD}. Kittel \cite{Kittel1948}, in 1949,  and  then Walker \cite{Walker1957}, in 1957, described discrete magnetostatic magnon modes in soft magnets such as yttrium iron garnet, when shaped into confined geometries. Recently magnons have grown in interest in hybrid quantum architectures \cite{Lachance-Quirion2019}, as potential quantum transducers, due to their ability to couple to  different types of quantum systems including microwave photons \cite{Tabuchi2014, Zhang2014}, superconducting qubits\cite{Tabuchi2015, Lachance-Quirion2017}, optical cavity modes \cite{Zhang2016B}, mechanical breathing modes \cite{Zhang2016}, nearby spins \cite{Fukami2024}, and in the case when the soft-magnet is trapped or levitated, coupling of the magnons to the center-of-mass motion\cite{Gonzalez-Ballestero2020, Kani2022, Xiong2025}.

Magnons are one of the systems that remain quantum even at room temperature. Their research began with the discovery of the Bloch law \cite{Bloch1930ZurTD}, which describes thermal demagnetization of magnets via spin-wave excitation. In studying spin waves in confined geometries, Kittel\cite{Kittel1948} and Walker\cite{Walker1957} discovered discrete magnonic modes, which are now named after them. These magnonic modes are particularly interesting for hybrid architectures\cite{Lachance-Quirion2019} as quantum transducers due to their ability to strongly couple to various quantum media: microwave photons\cite{Tabuchi2014, Zhang2014}, superconducting qubits\cite{Tabuchi2015, Lachance-Quirion2017}, optical cavity modes, breathing modes, spins, and potentially the center-of-mass motion\cite{Gonzalez-Ballestero2020, Xiong2025}. The collective nature of magnonic modes enables ultra-strong coupling, which was predicted to lead to size-independent cooling of levitated yttrium iron garnet (YIG) \cite{Kani2022}. The ability to cool size-independently and the potential to couple to nonlinear quantum systems are particularly interesting for studies of macroscopic quantum mechanics. A macroscopic YIG sphere was successfully levitated in a cryogenic environment  \cite{Fuwa2023}.  In this report, we demonstrate an alternative method of YIG levitation using a magnetic Paul trap (MPT) at room temperature. A magnetic Paul trap was used to levitate hard magnets up to a centimeter in size\cite{Sackett1993, Perdriat2023, Janse2024, janse2026chip}. There are a variety of designs for a magnetic Paul trap ranging from large setups using tightly wound coils and strong magnets \cite{Sackett1993},  through to mechanically rotating hard magnets \cite{Perdriat2023} and a chip scale integrated planar design\cite{Janse2024, janse2026chip, Perdriat2023}. For magnon-based applications, soft magnets are preferred over hard magnets. Therefore, we adapted a planar MPT for levitating a soft magnet, particularly YIG. In this letter, we present the stability analysis of the MPT in all 6 degrees of freedom, followed by our experiment on YIG levitation and characterization of its motion. In the final part, we estimate the enhanced coupling strength between the center-of-mass motion and magnons for experimentally feasible parameters, demonstrating strong cooperativity despite high dissipation.

\begin{figure}
\includegraphics[width=\columnwidth]{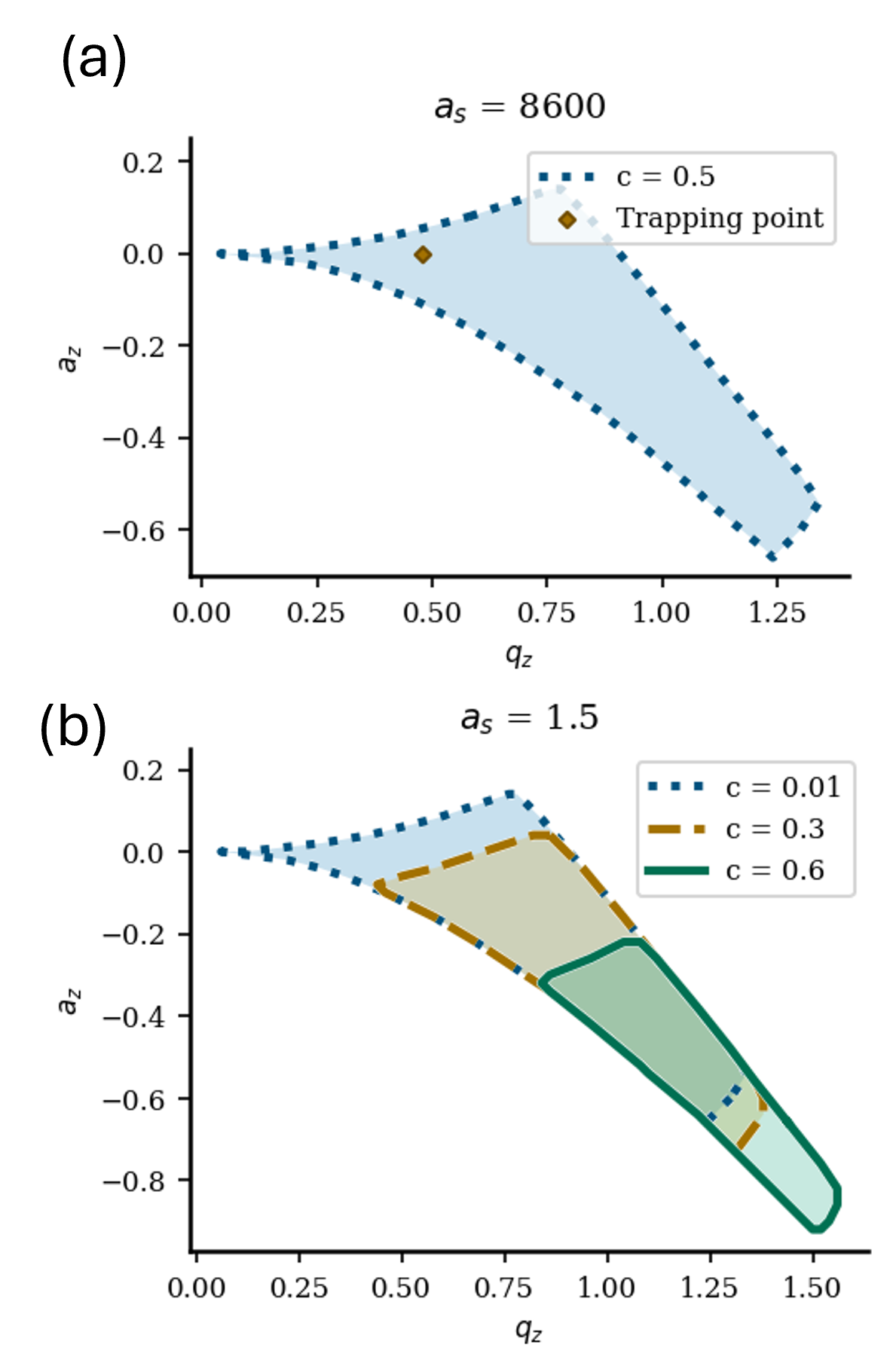}

\caption{\label{fig:stability} We study the conditions for stability of the levitation of the magnet in the magnetic Paul trap, so that we can understand the parameter values for the best trapping. We note that this stability analysis is more complex than for an ion trap, as the magnet has tilting degrees of freedom, which are not present in an ion. We plot the numerically estimated fundamental stability region on the [$a_z - q_z$] plane. $a_z$ ($q_z$) quantifies the strength of the static (time-dependent) trapping fields. We choose fixed values of $a_s$, which quantifies the librational dynamics, and fixed values for $c$, which quantifies the coupling strength between the librational and translational motion. $c$ can be modified by altering the AC field frequency and the magnetic gradient. In (a), we plot the stability diagram of the values we chose in the experiment. While in (b), we show how the stability region changes for different values of c. The brown diamond indicates the actual trapping parameters of the experiment.
 }
\end{figure}


Previously, a classical study of levitation of magnets in a MPT was conducted under the assumption of a magnet with constant magnetization and a fixed magnetization direction relative to the magnet's orientation \cite{Sackett1993, Perdriat2023}.  The types of magnets levitated in a MPT have only been hard magnets \cite{Sackett1993,Perdriat2023,Janse2024,janse2026chip}. We argue that, in the classical limit, a soft anisotropic magnet with saturated magnetization would also possess magnetization that is constant and fixed relative to orientation. By applying a strong external magnetic field $B_0$ exceeding the saturation magnetization of a soft magnet, we ensure that the magnitude of the magnetization vector remains constant. The direction of the magnetization relative to the orientation of the soft magnet will also be preserved as the anisotropic forces align the easy axis of the crystal along $\boldsymbol{B_0}$. Thus, by satisfying the condition $B_0 > B_s$, the classical theory developed to date for hard magnets in a MPT should also be valid for the levitation of soft magnets such as YIG. 



We analyze the stability using the model, where a magnetized body is under a total magnetic field $\bm{B} = \bm{B}_0 + \bm{B}_1 + \bm{B}_{2dc} + \bm{B}_{2ac}$ where $\bm{B}_0$, $\bm{B}_1$, $\bm{B}_{2dc}$ are static multipole magnetic fields eq.(\ref{equn:B0}-\ref{equn:B2dc}), $\bm{B}_{2ac}$ is an axisymetric AC curvature (Eq.\ref{equn:B2ac}).

\begin{subequations}
\begin{eqnarray}
\bm{B}_0 &=& B_0 \bm{e}_z, \label{equn:B0} \\
\bm{B}_1 &=& B_1^{'}\left( z \bm{e}_z - \frac{x}{2}\bm{e}_x - \frac{y}{2} \bm{e}_y \right), \label{equn:B1} \\
\bm{B}_{2dc} &=& \frac{B^{''}_{2dc}}{2}\bm{Q}_z
-\frac{B^{''}_{2dc}}{2}\bm{Q}_{xy}, \label{equn:B2dc}\\
\bm{B}_{2ac} &=& \frac{B^{''}_{2ac}}{2}\cos{\left(\Omega t\right)}  \bm {Q_z}
- \frac{B^{''}_{2ac}}{2} \cos{\left(\Omega t\right)} \bm{Q}_{xy}.  \label{equn:B2ac}
\end{eqnarray}
\end{subequations}

Here, $\Omega$ is the angular frequency of an AC field, $x, y, z$ are spatial coordinates, $\bm{Q}_{xy} =x z \bm{e}_x +y z \bm{e}_y$, $\bm{Q_z} = \left(z^2-\frac{(x^2+y^2)}{2}\right)\bm{e}_z$ and $\bm{e}_x, \bm{e}_y, \bm{e}_z$ are unit vectors of Cartesian basis.

We consider a case in which the magnetization of the magnet is dominated by a static, homogeneous field $B_0$. Meaning that $B_0 \gg B_1^{'} \Delta x_i$, $B_0 \gg B_{2dc}^{''} \Delta x_i^2$,  $B_0 \gg B_{2ac}^{''} \Delta x_i^2$. Here, $\Delta x_i$ is the displacement of a magnet relative to the center of the multipole fields $x = y = z = 0 $.

The magnet in a magnetic fields eqs.(\ref{equn:B0} - \ref{equn:B2ac}) in small-motion limit has equations of motion as follows: 

\begin{subequations}
\begin{eqnarray}
\frac{\partial^2 z}{\partial \xi^2} &+& (a_z - 2 q_z\cos{2\xi})z = 0 \label{equn:Mathieuz}\\
\frac{\partial^2 r}{\partial \xi^2} &+& (a_r - 2 q_r\cos{2\xi})r - c s = 0 \label{equn:Mathieur}\\
\frac{\partial^2 s}{\partial \xi^2} &+& a_s s - c r = 0 \label{equn:Mathieus}
\end{eqnarray}
\end{subequations}

Where $x = -y = r$, $s = \beta \sqrt{\frac{J}{m}} = \gamma \sqrt{\frac{J}{m}}$, $\alpha$, $\beta + \pi/2$, $\gamma$ are Euler angles in $\boldsymbol{zyz}$ convention, $\alpha = 0$ (see Supplementary Material).  $\xi = \Omega t / 2$, $q_z = - 2q_r = \frac{2 \mu B_{2ac}^{''}}{m \Omega^2}$, $a_z = - 2a_r = \frac{4 \mu B_{2dc}^{''}}{m \Omega^2}$, $a_s = \frac{4\mu B_0}{J\Omega^2}$, $c = \sqrt{10}\frac{\mu B_1^{'}}{m R \Omega^2}$. 

In the Supplementary Material, we provide a more detailed derivation of the equations (\ref{equn:Mathieuz} - \ref{equn:Mathieus}). The Eq. \ref{equn:Mathieuz} is an independent Mathieu equation. Thus, the stability of the z-mode is described by well-known Mathieu stability tongues\cite{Ince1927-xm}. Eq. \ref{equn:Mathieur} and \ref{equn:Mathieus} form a system of coupled equations, where Eq. \ref{equn:Mathieur} is the Mathieu equation, and Eq. \ref{equn:Mathieus} is the harmonic oscillator. This system we represent as the vector Mathieu equation \cite{Landa2012}: 

\begin{eqnarray}
    \frac{\partial^2\bm{u}}{\partial\xi^2} + \left( A - 2 Q \cos{2 \xi} \right) = 0
     \label{equn:MathieuVector}
\end{eqnarray}

Here $ \bm{u} = \begin{pmatrix} r \\ s \end{pmatrix}$, $A = \begin{pmatrix} a_r & -c \\ -c & a_s \end{pmatrix} $, $Q = \begin{pmatrix}
    q_r & 0 \\ 0 & 0
\end{pmatrix}$

Stability regions for this equation we will estimate using Floquet theory \cite{Folkers2018}. By substitution of the general form of the vector Mathieu equation solution \cite{Landa2012}, we obtain a matrix equation for characteristic exponents $\Xi_j$, $j=1,2$. 

\begin{eqnarray}
   M &=& A -\Xi_j^2 I - Q\left( \left(D_2 - \left( D_4 - \left(D_6 -...\right) \right)  \right)^{-1}  \right)^{-1}  \nonumber
    \\ &+& Q\left( \left(D_{-2} - \left( D_{-4} - \left(D_{-6} -...\right) \right)  \right)^{-1}  \right)^{-1} = 0 
    \label{equn:characteristicExponentMatrix}
\end{eqnarray}

Here $D_{2n} = Q^{-1}\left(A - \left( 2 n + \Xi_j \right)^2 I \right)$, $I$ is an identity matrix, $X^{-1}$ means inverse of a matrix $X$.

Eq.\ref{equn:characteristicExponentMatrix} has non-zero solutions only when $\det{M} = 0$. Thus, we find $\Xi_j$ by numerically solving this determinant equation. We solve equation $\det{M} = 0$ numerically for $\Xi_j$, and by checking the stability condition \cite{Landa2012} $\Im{\Xi} = 0$ for all found solutions, we find the stability region for coordinates $r$ and $s$ depending on $a_r$, $q_r$, $s$, and $a_s$. The intersection of the stable regions for the r, z, and s modes is the trap's stability region. We demonstrate the fundamental stable region of the trap in Fig. \ref{fig:stability}. We choose $[a_z - q_z]$ plane for fixed $s$ and $a_s$.  Subfigure a) shows the stability diagram for the parameters in our experiment. The brown diamond indicates the experiment's actual trapping parameters. The subfigure b) shows the parameters, where the stability region changes significantly with varying coupling. Stability regions for three chosen couplings $c$ are shown in different colors.


\begin{figure*}
\includegraphics[width=\linewidth]{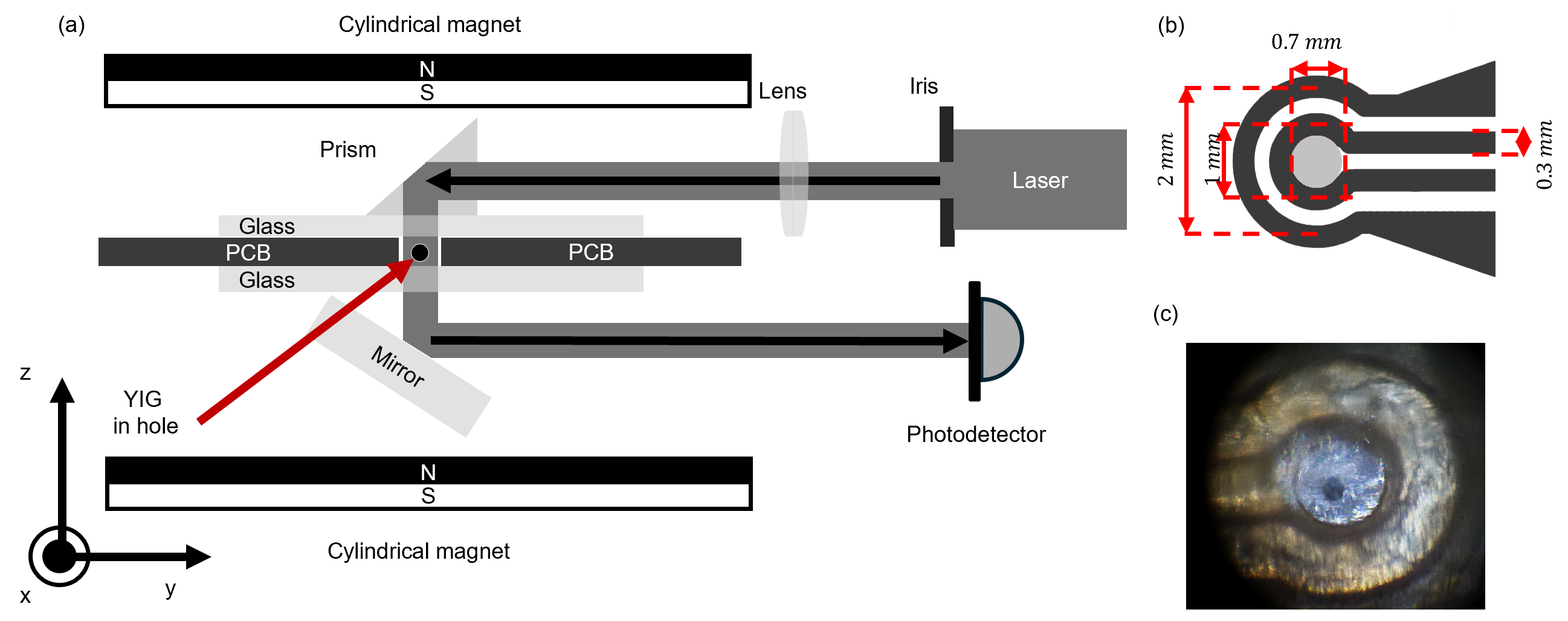}

\caption{\label{fig:experiment} Experimental setup for levitating a YIG sphere using a magnetic Paul trap and measuring motion using optical detection. a) We trap the YIG using a combination of a time-dependent magnetic field delivered by circulating currents flowing through electrodes on a printed circuit board and a static magnetic field generated by permanent magnets. The YIG is prevented from falling out from the trapping region by covering PCB hole with glass cover slides. We detect the motion by shining laser light down through the PCB hole and onto a photodiode. We use a lens to focus the light at the YIG's location, to amplify the motional signal at the photodetector.  b) Photo of loaded YIG in the MPT }
\end{figure*}


\vspace{\baselineskip}
The schematic diagram of the experimental setup is shown in Fig. \ref{fig:experiment}(a). We provide static magnetic components $\boldsymbol{B_0}$, $\boldsymbol{B_1}$, $\boldsymbol{B_{2dc}}$ using two  N52-grade cylindrical neodymium rare-earth disc magnets with a diameter of 2 inches and a thickness of 1/2 inch, provided by "KJ magnetics" in Helmholtz configuration. The separation between magnet surfaces is $37.6$ mm. The measured minimum magnetic field is $0.247$ T. We fabricated the trap on a custom-built PCB with two current electrodes, one with a radius of $1$ mm and the other $2$ mm. The thickness of the electrodes is $35$ um and the width is $0.3$ mm as shown in Fig. \ref{fig:experiment}(b). Inside the PCB, we drilled a hole $0.7$ mm in diameter to house the YIG. To the Top and bottom surfaces of the PCB, we glued thin microscope slide glass to prevent loss of the YIG. We control the magnetic gradient $B_{1}^{'}$ by offsetting the position of the trap along the z-axis. This, however, introduces undesired static curvature $\bm{B}_{2dc}$. The trapping AC magnetic curvature is provided by $\boldsymbol{B_{2ac}}$, which is generated by counter-circulating currents $I_{1ac} = 0.594, I_{2ac} = 1.223 A_{RMS}$ in the inner and outer electrode, respectively. The frequency of both currents is $140$ Hz. The static magnetic curvature $\boldsymbol{B_{2dc}}$ is compensated by biasing current electrodes with currents $I_{1dc} = 0.061, I_{2dc} = 0.111 A$. Using these parameters, we trapped a $0.2$ mm-diameter YIG from "Microsphere Inc." with a saturation magnetization $B_s$ = $0.178$ T. The image of a trapped YIG is shown in  Fig. \ref{fig:experiment}(c). The corresponding parameters we estimate as $q_z = 0.48$, $a_z \simeq 0$, $a_s \simeq 8600$, $c \simeq 0.5$, the stability plot of experimental parameters is shown on Fig. \ref{fig:stability}(a).

We measure levitated YIG motion by detecting backscattered light with a photodetector. As shown in Fig. \ref{fig:experiment}(a), we control the beam thickness by varying the iris diameter, focus the laser light onto the YIG sphere with a lens, and then detect the change of passed light with a photodetector. For analysis, we use the detector's voltage.

\begin{figure*}
\includegraphics[width=\linewidth]{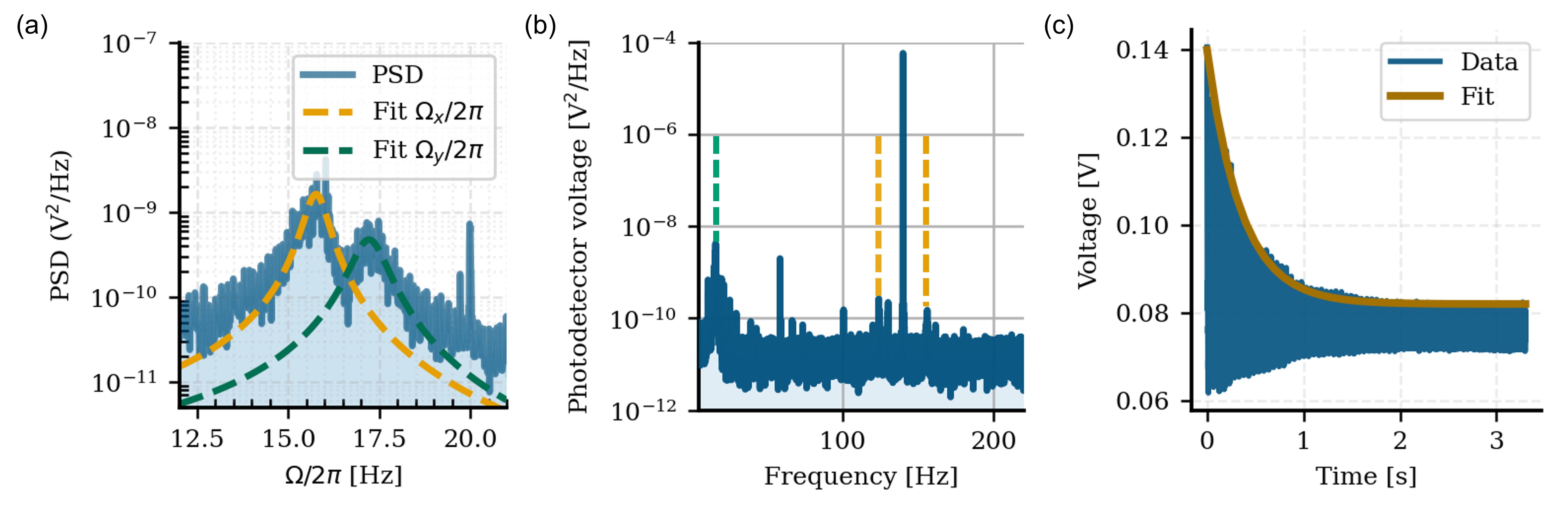}

\caption{\label{fig:experimentdata} Characterizing motion of the trapped in MPT YIG. We measure the motional PSD of trapped YIG in MPT with parameters  $I_{1ac} = 0.48$ $A_{RMS}$, $ I_{2ac} = 1$ $A_{RMS}$, $ I_{1dc} = 56$ mA, $ I_{2dc} = 113 $ mA, $ \Omega/2\pi = 140 Hz$. a) The secular modes of the translation motion. We fitted the data with a Lorentzian and obtained the frequencies $f_x = 15.8 Hz $ and  $f_y = 17.2 Hz$ along with Q-factors: $Q_x = 25, Q_y = 18$. b) Motional spectrum of trapped YIG. The spectrum consists of secular frequencies marked by a green dashed line, the micromotion mode at trap current frequency $f_m = 140 Hz$, and sidebands $f_m \pm f_x$ and $f_m \pm f_y$ that are marked by yellow dashed line. We conclude that these frequencies correspond to the x and y modes because they are closely spaced. Spectrum consists of other harmonics, which we do not associate with secular motion, as they are too narrow for a room pressure experiment. c) For trapping parameters $I_{1ac} = 0.587$ $A_{RMS}$ ,$ I_{2ac} = 1.175$ $A_{RMS}$, $  I_{1dc} = 59$ mA, $ I_{2dc} = 116$ mA, $\Omega/2\pi = 140$ Hz we performed the ring-down measurements. Firstly, we modulated the amplitudes of the trapping AC currents $I_1$ and $I_2$ with a modulation frequency $f_m = 19.3$ and a modulation depth of 1\%. After we switched off the modulation, we measured the decaying signal and fit it to an exponential function. From the fitted decay we calculated Q-factor of 29.  }
\end{figure*}

We characterize the PSD of the YIG sphere trapped with the following current amplitudes and biases $I_{1ac} = 0.48$ $A_{RMS}$, $ I_{2ac} = 1$ $A_{RMS}$, $ I_{1dc} = 56$ mA, $ I_{2dc} = 113 $ mA and frequency $ \Omega/2\pi = 140 Hz$. We show the PSD in the Fig. \ref{fig:experimentdata}(b). The spectrum shows two secular peaks at $f_x = 15.8$ Hz and $f_y = 17.2$ Hz. We also see the micro-motion frequency at $f_m = 140 Hz$ and sidebands $f_m \pm f_x$ and $f_m \pm f_y$. We conclude that these frequencies correspond to the x and y modes because they are closely separated. The observed motional spectrum contains other harmonics, which we do not associate with secular motion, as they are too narrow for a room pressure experiment. 



We fitted the peaks to a Lorentzian to obtain the Q factor and secular frequencies. Fig. \ref{fig:experimentdata}(a) shows the fit of the secular peaks.  We obtain $Q_x = 25$ and $Q_y = 18$.

Finally, we performed a ring-down measurement for the parameters $I_{1ac} = 0.587$ $A_{RMS}$, $ I_{2ac} = 1.175$ $A_{RMS}$, $  I_{1dc} = 59$ mA, $ I_{2dc} = 116$ mA, $\Omega/2\pi = 140$ Hz. We excite the YIG secular mode by modulating the amplitude of the trapping current with frequency $\Omega_m /2 \pi = 19.3$ Hz and modulation depth $1\%$. After we switch off the modulation, the YIG amplitude starts to decay, as shown in Fig.\ref{fig:experimentdata}. We fit the measured signal to an exponential function and calculate $Q = 31$. This value is close to the value obtained by fitting with a Lorentzian function.

\vspace{\baselineskip}
We now explore the potential of the quantum magno-mechanical system for a YIG sphere levitated in a magnetic field gradient in a harmonic trap with low motional damping. We will compute the cooperativity $C_{mb}$, which is a measure of how strongly the quantum information is exchanged between the magnon$\leftrightarrow$phonon (centre of mass-COM), in the presence of damping. We will see that in the case of strong magnon driving $C_{mb} > 1$, indicating strong interacting quantum dynamics. 

We follow previous works \cite{Kani2022, Xiong2025}, and consider a YIG sphere in bias magnetic field ${\bf B}=(B_0+\xi x){\bf e_z}$, with the Hamiltonian $\hat{H}_{tot}=\hat{H}_{mag}+\hat{H}_{mech}$, where the latter is the COM dynamics in the harmonic trap $2V(x)=m\omega_b^2\hat{x}^2$, and where $\hat{H}_{mag}$, involves a volume integration of the YIG's magnetic energy, which is a function of $x$, due to the magnetic field gradient. Making the standard approximations (Holstein Primakoff), one arrives at the magno-mechanical Hamiltonian:
\begin{gather}\label{quantized-hamiltonian-ladder}
    \hat{H} / \hbar = \omega_b \hat{b}^\dagger  \hat{b} + (\omega_0 + \Delta_l) \hat{m}^{\dagger} \hat{m} + g_{mb} (\hat{b}^{\dagger} +\hat{b}) \hat{m}^{\dagger}\hat{m}.
\end{gather}
where $\hat{b}(\hat{m})$, are the phonon(magnon) destruction operators, $\omega_0 = B_0 \gamma_e$ is the bare magnonic frequency, $\Delta_l = 2g_{mb}^2 N s/{\omega_b}$, is a slight magnon frequency shift due to the mechanical coupling, where $\gamma_s = N \gamma s$, $N$ is the number of spins in the YIG sphere, $\gamma_e$ is the gyromagnetic ratio, and $s$ is the spin number, and $\omega_b$ is the mechanical trap frequency. The single magnon magnomechanical coupling rate $g_{mb} = \xi \gamma x_{zpf}$, where $x_{zpf}=\sqrt{\hbar/(2m\omega_b)}$, is the width of the ground state wavefunction in the trap. 

\begin{figure}[t!]
\includegraphics[width=\columnwidth]{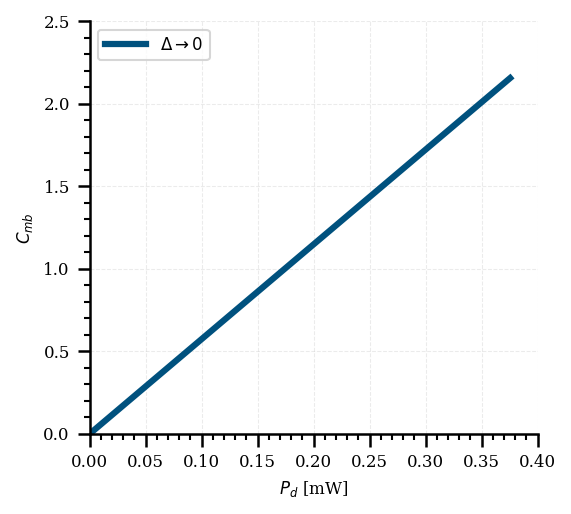}
\caption{\label{fig:fig4} Magnon enhanced cooperativity depending on the driving magnon power: We plot the cooperativity between the magnon and center of mass motion of levitated YIG for zero detuning. When $C_{mb}$ exceeds unity, the two modes are strongly coupled and quantum effects can be become dominant. 
}
\end{figure}

Including a magnon drive $\hat{H}_{MW}=+ i\Omega_d(\hat{m}^{\dagger}e^{-\omega_d t} - \hat{m}e^{i\omega_d t })$, where the Rabi frequency $\Omega_d = \frac{\sqrt{5}}{4}\gamma \sqrt{N} B_d=\gamma_e\sqrt{(5\mu_0\rho_sdP_d)/(12 c)}$, $B_d(P_d)$ is the amplitude(power) of the driving magnetic field, and $c$ is the speed of light. We can move to a frame rotating with the drive to find
 \begin{gather}\label{driven-rotating-frame-hamiltonian}
        \hat{H} / \hbar = \omega_b \hat{b}^\dagger  \hat{b} + \Delta \hat{m}^{\dagger} \hat{m} + g_{mb} (\hat{b}^{\dagger} +\hat{b}) \hat{m}^{\dagger}\hat{m} + i\Omega_d(\hat{m}^{\dagger} -\hat{m}).
    \end{gather}
where $\Delta=\omega_d-\omega_0$. From Eqs. (\ref{quantized-hamiltonian-ladder}, \ref{driven-rotating-frame-hamiltonian}), we arrive at  a type of optomechanical/magnomechanical coupling between the magnon occupation number and centre of mass motion.  Following the normal procedure in optomechanics, we assume that under strong magnon driving both the mechanics and magnon fields are driven to near stationary values with additional quantum perturbations and we set $\hat{m} = \Bar{m} + \delta\hat{m}$,  $\hat{b} = \Bar{b} + \delta\hat{b}$,  where $(\Bar{m},\Bar{b})$, are the steady state values of the modes under driving and damping and the mean magnon occupation is now $n=|\Bar{m}|^2$. One can show that these steady state values are solutions of the following equations:
\begin{equation}\label{equation-n-magnon-power}
\left(\kappa_m^2 + \Delta^2\right)n - \frac{4\Delta g_{mb}^2 n^2}{\omega_b} + \frac{4 g_{mb}^4 n^3}{\omega_b^2} = |\Omega_d|^2.
\end{equation}
One can find a new linearised optomechanical coupling for the quantum fluctuations $(\delta\hat{m},\delta\hat{b})$, (dropping the $\delta$s for clarity), to be:
\begin{gather}
    \hat{H}_{mb}/\hbar =  G_{mb} (\hat{m}^{\dagger}\hat{b}+\hat{m}\hat{b}^\dagger )\;\;,
\end{gather}
where the amplified coupling strength $G_{mb}=\sqrt{n}g_{mb}$, which is a function of the steady state magnon occupation $n$. We can now study the quantum cooperativity $C_{mb}=G_{mb}^2/(\kappa_m\kappa_b)$, where $\kappa_m(\kappa_b)$, are the damping rates of the magnons(phonons). For a sample evaluation we assume $\kappa_m/2\pi=1$ MHz, and $\kappa_b/2\pi\sim 0.4$ mHz, of a YIG sphere of diameter $d=0.2$ mm with $Q = 10^{3}$. Choosing a magnetic field gradient of  $\xi=252$ mT/m, spin density $\rho_s = 4.3 \times 10^{27}{\rm m}^{-3}$, gyromagnetic ratio $\gamma_e/2\pi = 28$ GHz/T, temperature of $T = 4$ K, saturation magnetization of YIG, $B_{sat} = 0.254$ T,  mechanical frequency $\omega_b/2\pi = 62$ Hz, results in the single-magnon coupling strength $g_{mb}/2\pi \sim 17$ $\mu$Hz. Solving Eqn (\ref{equation-n-magnon-power}), using these values for the magnon steady state occupation $n$, we plot the cooperativity $C_{mb}$ in Fig. \ref{fig:fig4}, as a function of drive power $P_d$, for zero detuning.  From Fig \ref{fig:fig4}, the setup  reaches moderately high cooperativity, despite the large magnon damping rate. We limit the maximum drive power to avoid breaking the Holstein-Primakoff approximation which requires only a small fraction of the spins should be excited in the entire YIG sphere. This large value of the cooperativity is a good indication that magno-mechanical centre of mass couplings can be used to explore strong coupling and perhaps use the mechanical mode for the long lived storage of quantum information.

In conclusion, we experimentally demonstrate the continuous trapping of a millimeter-sized YIG sphere with the aid of a magnetic Paul trap for at least an hour. Our experiment demonstrates the mechanical Q-factor $\sim 25$ and the mechanical secular frequencies $f_x = 15.8$ Hz and $f_y = 17.2$ Hz. We cannot measure z-mode due to the limited sensitivity of the chosen optical measurement setup along the z-axis. Furthermore, we theoretically analyze the stability of our model and observe that the experimental operating point resides within the stable regime. Our analysis demonstrates that coupling between the librational and center-of-mass modes changes the stability region. We then shift to a quantum description of the model, in which the magnetic gradient, initially designed to counteract gravity, facilitates the coupling between the quantized spin wave of the YIG sphere (magnon) and the center-of-mass motion of the YIG sphere. Furthermore, by driving the magnon, we demonstrate that the enhanced cooperativity can exceed unity, indicating a potential application in quantum information processing.

\vspace{\baselineskip}

We wish to acknowledge funding from the Okinawa Institute for Science and Technology for this research.\\[1em]

\section*{Author Declarations}
\subsection*{Conflict of Interest}
The authors have no conflicts to disclose.

\section*{Data Availability Statement}

The data supporting the findings presented in this paper are available from the corresponding author, AY, upon reasonable request.



\section*{References}
\putbib[aipsamp]
\end{bibunit}


\clearpage
\onecolumngrid

\begin{bibunit}
    
\setcounter{equation}{0}
\renewcommand{\theequation}{S\arabic{equation}}

\begin{center}
       \textbf{ \LARGE Supplementary material for Levitation of a YIG sphere using a magnetic Paul trap - towards strongly coupled quantum magno-mechanics}\\[1em]

    {\large
    Andrii Yakymenko$^{1}$, S. Das$^{1}$, and J. Twamley$^{1}$}\\[0.5em]
    \normalsize
    $^{1}$ Quantum Machines Unit, Okinawa Institute of Science and Technology, 1919-1 Tancha, Onna-son, Kunigami-gun, Okinawa, Japan 904-0495\\
 
\end{center}

\section*{Stability analysis}

We used expressions of Hamiltonian and magnetic moment provided in the Supplementary Material of previous work by Pedriat \textit{et al.}\cite{Perdriat2023}, to obtain the equations of motion of a magnet in the field $\bm{B}$ as follows:

\begin{subequations}
\begin{eqnarray}
    m\ddot{x} &=& -\frac{\mu B_{2ac}^{''}}{2}\cos{(\Omega t)} x -\frac{\mu B_{2dc}^{''}}{2} x \label{equn:EOMx} \\
    &-& \frac{\mu B_1^{'}}{2}\left(c_{\alpha}\beta + s_{\alpha}\gamma \right), \nonumber \\
    m \ddot{y} &=& -\frac{\mu B_{2ac}^{''}}{2}\cos{(\Omega t)} y -\frac{\mu B_{2dc}^{''}}{2} y \label{equn:EOMy} \\ 
    &-& \frac{\mu B_1^{'}}{2}\left(s_{\alpha}\beta - c_{\alpha}\gamma \right), \nonumber \\
    m\ddot{z} &=& -mg + \mu B_1^{'} + \mu B_{2ac}^{''} \cos{(\Omega t)} z + \mu B_{2dc}^{''} z \label{equn:EOMz} \\
    J \ddot{\alpha} &=& 0 \label{equn:EOMa}\\
    J \ddot{\beta} &=& -\mu B_0 \beta -\frac{\mu B_1^{'}}{2}(c_{\alpha}x + s_{\alpha}y) \label{equn:EOMb}\\
    J \ddot{\gamma} &=& -\mu B_0 \gamma -\frac{\mu B_1^{'}}{2}(s_{\alpha}x - c_{\alpha}y \label{equn:EOMg})
\end{eqnarray}
\end{subequations}

Here $x$, $y$, $z$ are CoM coordinates, $\alpha$, $\beta + \pi/2$, $\gamma$ are Euler angles in $\boldsymbol{zyz}$ convention. $c_{i}$ and $s_{i}$ are short notation for $\cos{(i)}$ and $\sin{(i)}$ respectively. The magnetic gradient $B_{1}^{'}$ has the primary purpose of applying a force to counter gravity, so we choose the $B_1^{'}$ satisfying the condition $mg = \mu B_1^{'}$.  Equation \ref{equn:EOMa} can be solved independently, giving the solution $\alpha =  \omega_\alpha t + \alpha_0$. In our and reported experiments \cite{ Perdriat2023, Janse2024, janse2026chip, Sackett1993}, the magnets under a homogeneous external magnetic field did not rotate. Therefore, we assume $\omega_\alpha = 0$. Due to axial symmetry, we choose a coordinate system where $\alpha_0 = 0$ without the loss of generality. We will simplify the system further by substituting $\tilde{y} = -y$. After plugging in $\alpha = 0$ and $y = \tilde{y}$, one will see that pairs of equations \ref{equn:EOMx}, \ref{equn:EOMb}, and \ref{equn:EOMy}, \ref{equn:EOMg} are equivalent. Therefore, it would be sufficient to solve only one pair.

Before writing down a new equations, we redefine spatial coordinates to match Mathieu form as follows: $x = -y = r$, and introduce a new variables $\xi = \Omega t / 2$ for time and for angles $s = \beta \sqrt{\frac{J}{m}} = \gamma \sqrt{\frac{J}{m}}$. Here, $J$ is a moment of inertia, $m$ is the mass of the trapped magnet. This new variables ensures that all angles and coordinates are expressed in units of length. For the spherical shape $\sqrt{\frac{J}{m}} = \frac{2}{5}R$ where R is a radius of a sphere. From now on, we consider a spherical trapped magnet. Aditionaly, we redefine physical parameters in dimensionless quantities $q_z = - 2q_r = \frac{2 \mu B_{2ac}^{''}}{m \Omega^2}$, $a_z = - 2a_r = \frac{4 \mu B_{2dc}^{''}}{m \Omega^2}$, $a_s = \frac{4\mu B_0}{J\Omega^2}$, $c = \sqrt{10}\frac{\mu B_1^{'}}{m R \Omega^2}$. This yields the equations of motion in the main text, which we write down in the vector form.

\begin{eqnarray}
    \frac{\partial^2\bm{u}}{\partial\xi^2} + \left( A - 2 Q \cos{2 \xi} \right) = 0
     \label{equn:MathieuVector}
\end{eqnarray}

Here $ \bm{u} = \begin{pmatrix} r \\ s \end{pmatrix}$, $A = \begin{pmatrix} a_r & -c \\ -c & a_s \end{pmatrix} $, $Q = \begin{pmatrix}
    q_r & 0 \\ 0 & 0
\end{pmatrix}$

The solution of the vector Mathieu equation is in the following form \cite{Landa2012}:

\begin{eqnarray}
    \bm{u}(\xi) &=& \sum_{j=1}^{2}\Bigl(  A_j e^{i \xi \Xi_j} \sum_{n=-\infty}^{\infty}\boldsymbol{C}^{(j)}_{2n}e^{i 2 n \xi}  \nonumber \\
     &+&  B_j e^{-i \xi \Xi_j} \sum_{n=-\infty}^{\infty}\boldsymbol{C}^{(j)}_{2n}e^{- i 2 n \xi} \Bigr)
     \label{equn:MathieuVectorSolution}
\end{eqnarray}

Here, index j corresponds to one of two partial solutions, $\Xi_j$ is a characteristic Mathieu exponent, $A_j$ are constants that depend on initial conditions, $C_{2n}^{(j)}$ are Fourier expansion coefficients.  According to Floquet stability theory \cite{Folkers2018}, the system will be stable when $\Im(\Xi_J) = 0$ for $j=1,2$.

Similarly to how it is done for the scalar Mathieu equation \cite{Leibfried2003}, by substituting the Eq. \ref{equn:MathieuVectorSolution} in Eq. \ref{equn:MathieuVector} we obtain the recursive relations for $\Xi$ as shown in the main text. 

\section*{References}
\putbib[aipsamp]
\end{bibunit}
\end{document}